\documentclass[twocolumn,aps,10pt]{revtex4-1}

\usepackage{amsmath}
\usepackage{amssymb}
\usepackage{amsfonts}
\usepackage{mathrsfs}
\usepackage{graphicx}

\newcommand{\inpng}[1]{#1}\newcommand{\ineps}[1]{}
\newcommand{\Dslash}{D\hspace{-6.5pt}\slash}

\renewcommand{\Re}{\text{Re}}

\newcommand{\LLL}{\mathscr{L}}
\newcommand{\HHH}{\mathscr{H}}

\newcommand{\E}{\text{E}}

\newcommand{\ket}[1]{|#1\rangle}
\newcommand{\braket}[2]{\langle#1|#2\rangle}
\newcommand{\HH}{\mathcal{H}}

\newcommand{\RR}{\mathcal{R}}
\newcommand{\dd}{\text{d}}

\newcommand{\ppp}{\\[3pt]}
\newcommand{\nppp}{\nonumber\\[3pt]}
\newcommand{\bppp}{\\[-9pt]}

\begin{document}

\title{Quantization of time and the big bang via scale-invariant loop gravity}

\author{Charles H.-T. Wang}

\author{Marcin Stankiewicz}

\affiliation{Department of Physics, University of Aberdeen, King's College, Aberdeen AB24 3UE, United Kingdom}

\begin{abstract}
We consider the background-independent quantization of a general scale-invariant theory of gravity with matter, which supports a conserved Weyl current recently suggested as a natural flow of time. For scalar-metric systems, a conformal class of Ashtekar-Barbero connection variables is constructed, which can be quantized using spin networks. Crucially, the quantum states become separable into the eigen states of the generator of the scale transformation and spin-network states in the Einstein frame. The eigen values consist of additional quantum numbers including a new type of fundamental frequency $\omega$ and energy $E = \hbar \omega$ with respect to a new local time $\tau$ carried by every spin-network vertex. The discretely distributed $\tau$ values as the ``quanta of time'' correspond to a functional time related to the integrated Weyl current in the classical theory. The Immirzi ambiguity of loop quantum geometry is removed by scale symmetry. To probe the quantum behaviour of the early Universe, the new formalism is applied to a scale-invariant homogenous and isotropic cosmological model coupled to a doublet of scalars with illustrative numerical simulations. The Einstein-frame volume is quantized in recently improved and regularized polymer representations with an arbitrary Immirzi parameter. The resulting unitary evolution of the quantum state of an expanding universe has a positive energy spectrum. A rescaling of the Immirzi parameter is equivalent to a translation in time without changing dynamics. The big bang can be identified in the past time limit when the expectation values of the Jordan-frame volume tend to zero. Remarkably, the quantized big bang is not replaced by a big bounce---a prevalent scenario in present loop quantum cosmology.
\end{abstract}

\maketitle

\section{Introduction}

There have been substantial recent interests in physics strongly influenced by an underlying scale invariance \cite{Blas2011, Bars2014, Salvio2015, Ferreira2016, Ferreira2017, Ferreira2018, Veraguth2017, Wang2018, Shaposhnikov2018, Wetterich2019}.
Here, we present significant new features of scale symmetry when it is amalgamated with a most fundamental description of nature, quantum gravity.

Time is ``famously lost'' in quantizing general relativity (GR), since general covariance prefers no temporal coordinates \cite{Isham1992}. Nonetheless, efforts continued to recover time from a nongravitational sector \cite{Rovelli1994, Husain2012}. GR is also nonrenormalizable perturbatively, which motivates loop quantum gravity (LQG) \cite{Thiemann2008, Rovelli2011, Ashtekar2017} as background-independent quantization using the polymer representations with countable kinematic states to avoid divergences. This leads to spin networks, the polymerlike fabrics of space, carrying quanta of areas and volumes.

The sizes of these ``chunks'' of geometry depend on an ambiguous Immirzi parameter $\gamma$ \cite{Immirzi1997a}. Phenomenological approaches exist e.g. to fix $\gamma$ so that the microstates of geometry around a black hole horizon reproduce the Bekenstein-Hawking entropy \cite{Perez2017}. Any fixed $\gamma$ predicts a minimum quantized volume which contributes to a widespread presumption that the big bang is replaced by a big bounce as demonstrated in many loop quantum cosmology (LQC) models \cite{Ashtekar2006, Ashtekar2008, Agullo2016}. Considering the universality of thermodynamics across different physical parameters and lack of experimental evidence of a particular $\gamma$, possibilities for new approaches to the Immirzi ambiguity remain open \cite{Rovelli2011}.

Here, we develop a unified approach to the problem of time, ambiguity of geometrical spectrum, and existence of the big bang in quantum gravity through scale symmetry. Scale invariance is found to be powerful in addressing these fundamental issues with profound implications.

\section{General scale-invariant systems}

In the Jordan frame, a general scale-invariant gravity-matter theory that encompass dilatonic and standard model-like interactions is described by the action
\begin{eqnarray}
S
&=&
\int
\sqrt{-g}
\,\Big\{
\frac{1}{2}
\RR \tilde{\phi}^2
+
\frac{1}{2}
(D\phi)^2
+
V(\phi)
\nppp&&
+
\frac14
F_{\mu\nu} F^{\mu\nu}
-
\Re\,
\psi^\dagger\, i \gamma^0
(\Dslash + \mu(\phi))\,\psi
\Big\}\,
\dd^4x
\label{LLLtot}
\end{eqnarray}
for the spacetime metric tensor $g_{\mu\nu}$ and $n_\phi$-dimensional scale field $\phi$, spinor fields $\psi$, and gauge connections $A_\mu$ with field strengths  $F_{\mu\nu}$.
To avoid a negative conformal factor relating $S$ to the Einstein frame, we have used an overall minus sign Eq. \eqref{LLLtot} without changing dynamics.
The metric has signature $(-,+,+,+)$, scalar curvature $\RR$, and determinant $g$. We adopt units where $c=\hbar=8\pi G = 1$.
The nonminimal gravitational coupling with the scalar is provided by the factor $\tilde{\phi}^2 = \xi_{ij} \phi_i \phi_j$ with a symmetric matrix $\xi$. A more restricted diagonal $\xi$ has previously been considered in \cite{Wang2018}.
The scalar potential $V(\phi)$ is a fourth order homogeneous polynomial in $\phi_i$ and can give rise to an effective cosmological constant, Higgs-type mechanisms, cosmic inflation, and hierarchy generation in fundamental couplings \cite{Ferreira2016}.

The gauge fields $A_\mu$ couple in a standard manner through covariant derivatives $D_\mu$ to the scalars with $(D\phi)^2 = D_\mu\phi_i\,D^\mu\phi_i$ and to the spinors with $\Dslash = \gamma^I e^\mu{}_I D_\mu$.
Here the Dirac matrices $\gamma^I$ satisfying the Clifford algebra
$\gamma^{(I}\gamma^{J)} = \eta^{IJ}$
and Hermiticity condition
$\gamma^I{}^\dag = \gamma^0\gamma^I\gamma^0$,
with a Yukawa coupling matrix
$\mu(\phi)$ homogeneous linearly in $\phi_i$ \cite{Bekenstein1980}.

Under a conformal, i.e. local rescaling, transformation with an arbitrary positive spacetime function $\Omega(x^\mu)$, we have
\begin{subequations}
\begin{eqnarray}
g_{\mu\nu}
\to
\Omega^2 g_{\mu\nu}
,\quad
\phi
\to
\Omega^{-1}\phi
,
\label{scltrans1}
\ppp
A_\mu
\to
A_\mu
,\quad
\psi
\to
\Omega^{-3/2}\psi .
\label{scltrans2}
\end{eqnarray}
\label{scltrans}
\bppp
\end{subequations}
The scale invariance of the action \eqref{LLLtot} means it remains unchanged under the simultaneous transformations \eqref{scltrans},
where $\Omega$ is an arbitrary positive constant in general, or spacetime function in the special case with $\xi_{ij}=\delta_{ij}/6$ and scale invariance becomes conformal invariance.

The scale invariance of the theory implies the existence of a conserved Noether current, which can be identified from the boundary terms of the on-shell variations of the Lagrangian density of action \eqref{LLLtot} under an infinitesimal scale transformation \eqref{scltrans}.
As generalized from Ref. \cite{Wang2018} with an arbitrary $\xi$ matrix, this yields the wave equation
\begin{eqnarray}
\Box
\phi^2
&=&
0
\label{seqc}
\end{eqnarray}
where $\phi^2 = \phi_i\phi_i$ and $\Box$ is the box (Laplace-Beltrami) operator using the curved spacetime metric.
Eq. \eqref{seqc} constitutes a conservation law for the current $\partial_\mu\phi^2$ corresponding to the Weyl scaling symmetry, which will therefore be referred to as the ``Weyl current.''
It provides a flow of harmonic time $\phi^2$ when $\partial_\mu\phi^2$ is timelike \cite{Wang2018}.

\section{Scale-invariant loop quantum gravity}

To focus on quantum gravity with scale invariance, we now consider the metric-scalar sector of the general theory described by action \eqref{LLLtot} with a zero scalar potential. We start the canonical analysis by adapting the Arnowitt-Deser-Misner (ADM) formulation to the geometrodynamics of the resulting system and split the spacetime metric $g_{\mu\nu}$ into the spatial metric $h_{ab}$, lapse function $N$, and shift vector $N^a$. As in GR, the spatial metric develops in time according to
$
\partial_t {h}_{ab}
=
N_{a;b} + N_{b;a}
-
2 N k_{ab}
$
where $k_{ab}$ is the extrinsic curvature tensor.

Here, the nonminimal coupling with scalars introduces additional terms in the spacetime decomposed Lagrangian density of the metric-scalar sector of action \eqref{LLLtot}, which we obtain after some algebras, to be
\begin{eqnarray}
\LLL
&=&
\frac{N \sqrt{h}}{2}
(k_{ab}k^{ab}-k^2+R)
\tilde{\phi}^2
-
N\sqrt{h}\,\Delta\tilde{\phi}^2
\nppp&&
+
\sqrt{h}\,k d_t\tilde{\phi}^2
+
\frac{\sqrt{h}}{2 N}
(d_t\phi)^2
-
\frac{N \sqrt{h}}{2}
(\partial\phi)^2
\label{eqLx}
\end{eqnarray}
up to a total divergence.
Above,
$k$ is the metric trace of $k_{ab}$,
$d_t = \partial_t - N^a \partial_a$ and
$(\partial\phi)^2 = \partial_a\phi_i\,\partial^a\phi_i$, with
$h$, $R$, $\nabla_a$, and $\Delta$ being the determinant, scalar curvature, metric connection, and the Laplace-Beltrami operator of $h_{ab}$ respectively.

At the level of the geometrodynamics, the canonical momenta of the metric and scalar follow as
\begin{eqnarray}
p^{ab}
&=&
-
\frac{\sqrt{h}}{2}(k^{ab}
-
h^{ab}k)\,\tilde{\phi}^2
-
\frac{\sqrt{h}}{2N}\,h^{ab}\,d_t\tilde{\phi}^2
\label{ppab}
\ppp
\pi
&=&
\frac{\sqrt{h}}{N}\,d_t\phi
+
2\sqrt{h}\,k\, \xi{\phi}
\label{piphi1}
\end{eqnarray}
respectively. These relations allow us to derive the corresponding Hamiltonian density of the metric-scalar system by eliminate the extrinsic curvature tensor from Eq. \eqref{eqLx} to be
$\HHH
=
N \HH + N^a\HH_a
$
in a completely constrained form as per Dirac's extended canonical theory, in terms of the diffeomorphism constraint
\begin{eqnarray}
\HH_a
&=&
\pi\partial_a\phi
-
2\,{\nabla}_b\, p^{b}{}_{a}
\label{Cdiffx}
\end{eqnarray}
and the Hamiltonian constraint
\begin{eqnarray}
\HH
&=&
\frac{2(3p_{ab}p^{ab}-p^2)}
{3\sqrt{h}\,\tilde{\phi}^2}
-
\frac{\sqrt{h}}{2}\,R\,\tilde{\phi}^2
+
\sqrt{h}\,\Delta\tilde{\phi}^2
\nppp&&
+
\frac{\sqrt{h}}{2}
(\partial\phi)^2
+
\frac{\pi^2}{2\sqrt{h}}
-
\frac{(p + 3 \xi{\phi}\,\pi)^2}
{3\sqrt{h}\,\xi\phi(\phi
+
6 \xi{\phi})}
\label{HHa}
\end{eqnarray}
where
$\pi^2 = \pi_i \pi_i$.

For the above metric-scalar system to be quantized using background independent loop representations, it is necessary to introduce SU(2) connection-type canonical variables. These can be constructed by extending the conformal canonical techniques recently reported in Ref. \cite{Wang2018}, resulting in the following ``conformal spin connection variables''
\begin{eqnarray}
A_a^i
&=&
\Gamma_a^i + \gamma K_a^i
\label{AAai}
\end{eqnarray}
in terms of the ``conformal extrinsic curvature''
$
K_a^i
=
h^{-1} (2 p_{ab} - h_{ab} p) E^b_i
$
where $E_a^i$ is the densitized triad of the spatial metric $h_{ab}$,
$E = \det E_a^i$, and associated extrinsic curvature $K_a^i$  \cite{Wang2018}.
The canonical momenta conjugate to connection \eqref{AAai} are given by $E^a_i/\gamma$.

The generator of conformal transformations \cite{Wang2005a, Wang2005b}, i.e. local rescaling, is given by
\begin{eqnarray}
C
&=&
K + \frac{1}{2}\phi\,\pi
\label{Sgen}
\end{eqnarray}
where $\phi\,\pi = \phi_i\pi_i$ and
$K=K_a^i E^a_i$
is a conformally generalized Thiemann complexifier density \cite{Thiemann2008}.

In terms of the connection variables \eqref{AAai}, we can recast the Hamiltonian constraint \eqref{HHa} into the form
\begin{eqnarray}
\HH
&=&
\tilde{\phi}^2\,\HH_\E
-
\frac{1 + \gamma^2\tilde{\phi}^4}
{\sqrt{E}\,\tilde{\phi}^2}
K_{[a}^{i} K_{b]}^{j}
E^a_i E^b_j
+
\sqrt{E}\,\Delta\tilde{\phi}^2
+
\frac{\pi^2}{2\sqrt{E}}
\nppp&&
+
\frac{\sqrt{E}}{2}
(\partial\phi)^2
+
\frac{K^2}{3\sqrt{E}\,\tilde{\phi}^2}
-
\frac{(K - 3 \xi{\phi}\pi)^2}
{3\sqrt{E}\, \xi{\phi}(\phi - 6 \xi{\phi})} .
\label{HHHa}
\end{eqnarray}
The generator of scale transformations, i.e. global rescalings, is given by
\begin{eqnarray}
\Pi
&=&
\int_\Sigma C\, \dd^3x
\label{sclgen}
\end{eqnarray}
so that relations \eqref{scltrans1} are recovered by e.g.
$\{\Pi,\sqrt{E}\}=\frac{3}{2}\sqrt{E}$
and
$\{\Pi,\sqrt{E}\}=-\frac{1}{2}\sqrt{E}$.
From Eq. \eqref{HHHa}, we also have $\{\Pi,\HH\}=-\frac{1}{2}\HH$.
Consequently, we can introduce a scale-invariant Hamiltonian constraint
\begin{eqnarray}
\bar{\HH}
&=&
\tilde{\phi}^2\sqrt{E}\,\HH
\label{SHH}
\end{eqnarray}
satisfying
$\{\Pi,\bar{\HH}\}=0$.

Using a conformal transformation with $\Omega=\tilde{\phi}(x)$, we obtain the conformally invariant Einstein-frame densitized triad
%
\begin{eqnarray}
\bar{E}_i^a
&=&
\tilde{\phi}^2 E^i_a
\ppp
\bar{A}_a^i
&=&
\bar{\Gamma}_a^i
+
\gamma\bar{K}_a^i
\end{eqnarray}
%
where $\bar{\Gamma}_a^i$ is the LC spin connection of $\bar{E}_i^a$ and
$\bar{K}_a^i = \tilde{\phi}^{-2}K_a^i$ is the conformally invariant Einstein-frame extrinsic curvature.

\begin{figure}
\inpng{\includegraphics[width=0.75\linewidth]{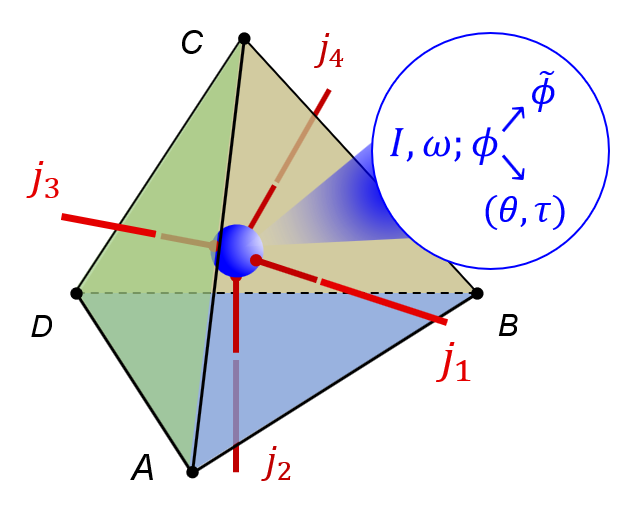}}
\ineps{\includegraphics[width=0.75\linewidth]{ScalarSpin.eps}}
\caption{New structure of a scalar-spin network:
A spin network vertex (blue) inside a tetrahedron with corners $A$, $B$, $C$, and $D$ is illustrated. Each of the four triangular faces is pierced by an edge (red) connected to the vertex, carrying spin quantum numbers $j_1$, $j_2$, $j_3$, and $j_4$. In addition to the intertwiner quantum number $I$, the vertex carries a new frequency quantum number $\omega$.
The vertex also carries the $n_\phi$-fold scalar $\phi$ which contains a local time $\tau$, as well as an SO($n_\phi$)-valued $\theta$, and contributes $e^{-i\omega \tau}$ to the quantum state of the scalar-spin network. The scalar $\phi$ also yields $\tilde{\phi}$ so that the Jordan-frame areas of the triangular faces scale with $\gamma/\tilde{\phi}^2$. The Jordan-frame volume of the space around the vertex scales with $\gamma^{3/2}/\tilde{\phi}^3$.
The scaling with $\tilde{\phi}$ is a feature of the present theory. Note that the Immirzi parameter $\gamma$ can be absorbed by rescaling $\tilde{\phi} \to \sqrt{\gamma}\,\tilde{\phi}$ with corresponding rescaling of other fields.
}
\label{fig:ScalarSpin}
\end{figure}

Although the quantized Einstein-frame geometric quantities depends on $\gamma$, the corresponding geometric quantities are scaled by a compensating power of $\tilde{\phi}$ as illustrated in Fig. \ref{fig:ScalarSpin}. The scale-invariance of the theory means any $\gamma$ can be absorbed by a global rescaling and therefore the theory has no Immirzi ambiguity.
The total Hilbert space is complete with the Hilbert space for the rotational degrees of freedom in the space of the scalar fields $\phi$.
This leads to a natural set of new canonical variables, namely
$(\bar{A}_a^i, \bar{E}_i^a)$,
$(\theta,L)$,
and $(\tau,C)$
with Poisson bracket relations including
\begin{eqnarray*}
\{\bar{A}_a^i(x), \bar{E}_j^b(x')\}
&=&
\gamma\,\delta^i_j\delta^b_a\delta(x,x')
\ppp
\{L_{ij}(x),L_{kl}(x')\}
&=&
2\delta_{i[l}L_{k]j}(x)\delta(x,x')
-
(i\leftrightarrow j)
\ppp
\{\tau(x),C(x')\}
&=&
-\delta(x,x')
\end{eqnarray*}
where we have introduced the SO($n_\phi$)-valued angles of rotation $\theta$ with antisymmetric angular momentum  tensor
$L = 2\phi\wedge\pi$
in the scalar space, i.e.
$L_{i j} = \phi_i \pi_j - \phi_j \pi_i$.
Furthermore, we have introduced $\tau$ by
\begin{eqnarray}
e^\tau
&=&
\phi^{-2}
\label{etau}
\end{eqnarray}
and will demonstrate its fundamental role as time in what follows.

Assuming both $\Pi$ and $\bar{\HH}$ are quantized into Hermitian operators in some Hilbert space and $[\Pi,\bar{\HH}]=0$ is satisfied. Then solutions of
$\bar{\HH}\Psi=0$ can be simultaneously the eigen states of $\Pi$
or their linear combinations. Using
\begin{eqnarray}
\Pi
&\to&
i\int_\Sigma \frac{\delta}{\delta\tau(x)}\, \dd^3x
\label{qsclgen}
\end{eqnarray}
the eigen states of $\Pi$ with some real eigen value $\omega$ have the separable form
\begin{eqnarray*}
\Psi
&=&
\Psi_\omega[\bar{A}_a^i, \theta]\, e^{-iW[\tau]}
\end{eqnarray*}
for a functional $W[\tau]$ linear in $\tau(x)$ so that
\begin{eqnarray*}
\omega
&=&
\int_\Sigma
\frac{\delta W}{\delta\tau(x)}
\,\dd^3x .
\end{eqnarray*}
%


Since the Hamiltonian constraint acts on the vertices of spin networks \cite{Ashtekar2003}, it is sufficient to consider $W[\tau]$ to have support at vertices $x_i$, and assign each vertex an additional quantum number $\omega_v$ so that
\begin{eqnarray*}
W[\tau]
&=&
\omega_v\, \tau_v
:=
\omega_v\, \tau(x_v)
\end{eqnarray*}
with
$\omega=\sum_v \omega_v$.
The quantum states are therefore expressed as superpositions of
\begin{eqnarray}
\Psi
&=&
\Psi_\omega(h(e_l)[\bar{A}],\theta(x_j))\,
e^{-i\omega_v\, \tau_v}
\label{Psiw}
\end{eqnarray}
using the cylindrical functions of $\bar{A}_a^i$,
with holonomies $h(e_l)[\bar{A}]$ over edges
$e_l$.

As an interesting example, in the case of
$
\xi_{i j}
=
\xi\delta_{i j}
$
with any nonconformal value of $\xi > 1/6$,
Eq. \eqref{HHHa} simplifies significantly allowing Eq. \eqref{SHH} to yield
\begin{eqnarray}
\bar{\HH}
&=&
\xi^2{\phi}^4\sqrt{E}\,\HH_\E
-
\big(1 + \gamma^2\xi^2{\phi}^4\big)
K_{[a}^{i} K_{b]}^{j}
E^a_i E^b_j
\nppp&&
+
\xi^2 {\phi}^2 \Delta{\phi}^2 E
-
\frac{\xi L^2}{2}
+
\frac{
2\xi C^2
}
{
6\xi-1
}
\label{SHH2}
\end{eqnarray}
where
$
L^2
= \sum L_{i j}^2
=\phi^2\pi^2-(\pi\phi)^2
$
is the squared total angular momentum in the scalar space.

The quantum constraint equation $\bar{\HH}\Psi=0$ using Eq. \eqref{SHH2} then yields the following functional Schr\"odinger equation
\begin{eqnarray}
i \frac{\delta}{\delta\tau(x)} \Psi
&=&
h(x) \Psi
\label{scheq}
\end{eqnarray}
where
%
\begin{eqnarray*}
h^2
&=&
\frac{6\xi-1}{4}\,L^2
-
\frac{6\xi-1}{2\xi}
\Big\{
\xi^2{\phi}^4\sqrt{E}\,\HH_\E
\nppp&&
-
\big(1 + \gamma^2\xi^2{\phi}^4\big)
K_{[a}^{i} K_{b]}^{j}
E^a_i E^b_j
+
\xi^2 {\phi}^2 \Delta{\phi}^2 E
\Big\} .
\end{eqnarray*}

From Eqs. \eqref{Psiw} and \eqref{scheq} we have
\begin{eqnarray}
h(x) \Psi_\omega
&=&
\omega_v\,
\delta(x_v,x)
\Psi_\omega
\end{eqnarray}
analogous to the time-independent Schr\"odinger equations.

If $\Sigma$ is chosen to be a constant time hypersurface ($\tau=$ spatial const.), then Eq. \eqref{qsclgen} reduces to
$
\Pi
\to
i {\partial}/{\partial\tau}
$
and so the functional Schr\"odinger equation implies the effective Schr\"odinger equation
\begin{eqnarray}
i \frac{\partial}{\partial\tau} \Psi
&=&
\int_\Sigma h\, \dd^3x\, \Psi .
\label{scheq2}
\end{eqnarray}

\section{Scale-invariant cosmological model}

As a first application to cosmology, let us consider two scalar fields $\phi_1$ and $\phi_2$ with momenta $\pi_1$ and $\pi_2$ and equal curvature coupling
$\xi_1 = \xi_2 = \xi$ for $1/6 < \xi < \infty$. In a spatially flat 
Friedmann model, the metric sector is described by a fiducial cell with unity coordinate volume with the Einstein-frame volume
$\bar{V} = \sqrt{\bar{E}}$. The corresponding Jordan-frame volume, which interacts directly with matter \cite{Esposito2001}, is
$V = \xi^{-3/2}\phi^{-3}\bar{V}$, where
$\phi=\sqrt{\phi_1^2+\phi_2^2}\,$.
Then $\bar{\HH}$ in \eqref{SHH2} reduces to
\begin{eqnarray}
\bar{\HH}
&=&
-
\frac{1}{3}\bar{K}^2
+
\frac{2\xi}
{6\xi-1}\,
\Pi^2
-
\frac{\xi}
{2}\,L^2
\label{hhhc}
\end{eqnarray}
with the canonical pairs of variables:
$(\bar{V},
\bar{B} = -\frac{2}{3}\bar{K}/\bar{V})$,
$(\theta = \tan^{-1} {\phi_2}/{\phi_1},
L = \phi_1 \pi_2 - \phi_2 \pi_1)$,
and
$(\tau = \ln\phi^{-2},-\Pi)$, where $\bar{K} = \bar{K}_a^i \bar{E}^a_i$
is the Einstein-frame mean extrinsic curvature.

In terms of canonical variables $\bar{V}$ and $\bar{B}$, the Einstein-frame
mean extrinsic curvature
$\bar{K} = -\frac{3}{2}\bar{B}\bar{V}$,
fiducial cell volume $\bar{V}$,
and gravitational Hamiltonian constraint
$\bar{H} = -\frac{3}{4}\bar{B}^2\bar{V}$, the algebra
\begin{eqnarray}
\{
\bar{V},\bar{H}
\}
=
\bar{K}
,\;
\{
\bar{K},\bar{V}
\}
=
\frac{3}{2}\,
\bar{V}
,\;
\{
\bar{K},\bar{H}
\}
=
-
\frac{3}{2}\,
\bar{H}
\label{PBKVH}
\end{eqnarray}
ensures that $\bar{K}$ is the generator of the metric sector scale transformations \cite{Thiemann2008, Achour2017, Achour2019}.

Since the Einstein-frame quantities in the metric sector have the same forms as in GR \cite{Wagoner1970}, we can express the Einstein-frame volume of the fiducial cell and its momentum in terms of the variables $\nu$ and $b$ satisfying $\{\nu,b\} = 2$ commonly used in LQC \cite{Agullo2016} as follows
\begin{eqnarray}
\bar{V} = \frac{\gamma\,\nu}{4}
,\quad
\bar{B} = \frac{2 b}{\gamma} .
\end{eqnarray}

The above variables are quantized in polymer representations, where the Hilbert space consists of dense states $\ket{\nu}$ diagonalizing $\nu$, with inner product $\braket{\nu}{\nu'}=\delta_{\nu\nu'}$ and action
$
{e^{-i \lambda b}} \ket{\nu} = \ket{\nu+2\lambda}
$
of exponentiated $b$ for any real $\lambda$.
The standard approach \cite{Ashtekar2006} corresponds to the regularization
$
b
\stackrel{\text{reg.}}{\longrightarrow}
\lambda^{-1}{{\sin\lambda b}}
$
of $b$ in the Hamiltonian constraint operator, where
$\lambda^2 = 4\sqrt{3}\,\pi\gamma\,\ell_\text{P}^2$ and $\ell_\text{P}$ is the Planck length.
It leads to a lattice of the basis states $\ket{\nu}$ with equal spacing $\Delta\nu = 4\lambda$.
However, this regularization dose not preserve the algebra \eqref{PBKVH}. To ensure $\bar{K}$ as the generator of the scale transformation in the metric sector as part of our scale-invariant quantization, we adopt a modified regularization
\begin{eqnarray}
b
&\stackrel{\text{reg.}}{\longrightarrow}&
\frac{\tan\lambda b}{\lambda},
\quad
\nu
\stackrel{\text{reg.}}{\longrightarrow}
\nu\,\cos^2\lambda b
\end{eqnarray}
on both $b$ and $v$ that preserves the algebra \eqref{PBKVH} recently suggested in Refs. \cite{Achour2017, Achour2019}.

We also employ symmetric factor orderings of $\nu$ and $b$ in quantizing $\bar{K}$, $\bar{V}$, and $\bar{H}$ into Hermitian operators, with their  regularized constructions and actions explicitly given by
\begin{subequations}
\begin{eqnarray}
\bar{K}\ket{\nu}
&=&
-\frac{3}{16 \lambda}\,
\big[
{\sin 2\lambda b}\,
{\nu}
+
{\nu}\,
{\sin 2\lambda b}
\big]
\ket{\nu}
\nppp
&=&
\frac{3i}{16 \lambda}\,
\big[
(\nu - 2\lambda)\ket{\nu - 4\lambda}
\nppp&&
-
(\nu + 2\lambda)\ket{\nu + 4 \lambda}
\big],
\label{KVHact1}
\ppp
\bar{V}\ket{\nu}
&=&
\frac{\gamma}{4}\,
{\cos \lambda b}\,
{\nu}\,
{\cos \lambda b}\,
\ket{\nu}
\nppp
&=&
\frac{\gamma}{16}\,
\big[
(\nu-2\lambda)\ket{\nu - 4\lambda}
+ 2\nu\ket{\nu}
\nppp&&
+
(\nu +2\lambda)\ket{\nu+4\lambda}
\big],
\label{KVHact2}
\ppp
\bar{H}\ket{\nu}
&=&
-\frac{3}{4\lambda^2 \gamma}\,
{\sin \lambda b}\,
{\nu}\,
{\sin \lambda b}\,
\ket{\nu}
\nppp
&=&
\frac{3}{16\lambda^2 \gamma}\,
\big[
(\nu-2\lambda)\ket{\nu - 4\lambda}
- 2\nu\ket{\nu}
\nppp&&
+
(\nu +2\lambda)\ket{\nu+4\lambda}
\big].
\label{KVHact3}
\end{eqnarray}
\label{KVHact}
\bppp
\end{subequations}

\begin{widetext}

\begin{figure}
\inpng{\includegraphics[width=0.9\linewidth]{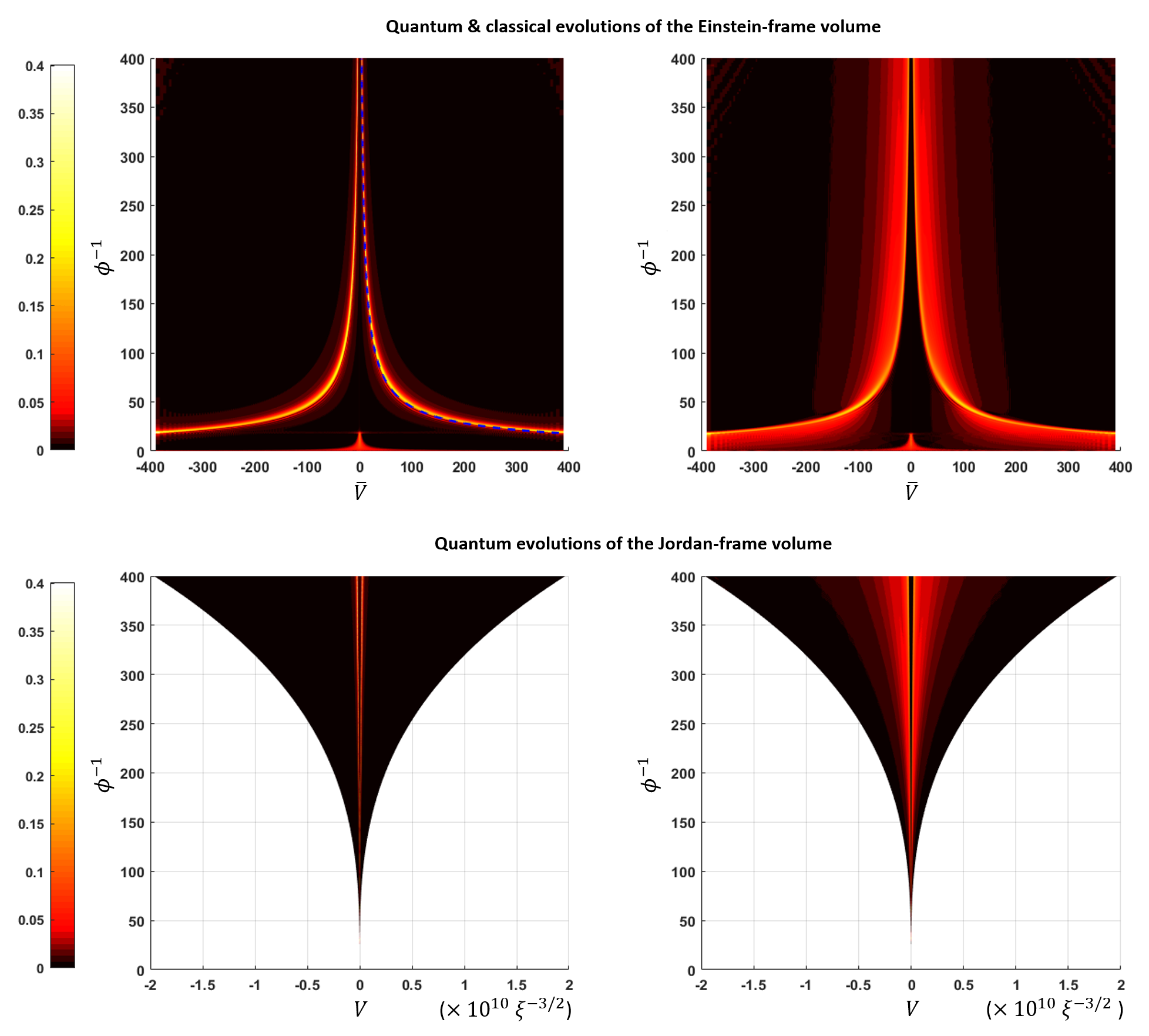}}
\ineps{\includegraphics[width=0.9\linewidth]{VolEinJor.eps}}
\caption{Numerical simulations of an expanding universe: The modulus squared of a parity-preserving wave function $\Psi(\nu)=\Psi(-\nu)$ is evaluated on a symmetric lattice with $\nu=4 \lambda (n-1/2)$ which excludes $\nu=0$, for a truncated range $1-n_{\max} \le n \le n_{\max}$ of $n$. The wave function evolves unitarily along the inverse scalar $\phi^{-1}=e^{\tau/2}$ shown vertically. Choosing $\gamma=1$, $\xi=2/9$, and truncation size $n_{\max} = 400$, an initial wave packet
at $\phi^{-1}=50$ corresponding to $\tau=-7.82$ shown in Fig. \ref{fig:Init} with $m=0$ leads to a nondispersive evolution of the probability distribution of the Einstein-frame volume (upper-left), where the corresponding classical solution (blue dash) follows closely the positive branch of the wave packet motion. For the same parameter and initial wave packet with $m=2$, the evolution becomes dispersive (upper-right). The corresponding quantum evolutions of the Jordan-frame volume, to which matter directly couples, are show in the lower-left plot with $m=0$ and in the lower-right plot with $m=2$, where the cosmic expansion is evident. Furthermore, the big bang can be identified at $\phi^{-1}\to 0$ i.e. $\tau\to-\infty$ where $V\to0$.  Similar simulations are obtained with an alternative symmetric lattice $\nu=4 \lambda n$ for $-n_{\max} \le n \le n_{\max}$ including $\nu=0$. Here, the choices of $m=0$ and $m=2$ are representative of nondispersive and dispersive volume evolutions respectively. The latter can be obtained with any nonzero $m$ values. A small ``lump'' appearing at the bottoms of upper plots is caused by the finite grid size, which can be reduced by increasing $n_{\max}$ as shown in Fig. \ref{fig:AbsVolEin}. }
\label{fig:Vol}
\end{figure}

\end{widetext}

\begin{figure}
\inpng{\includegraphics[width=1\linewidth]{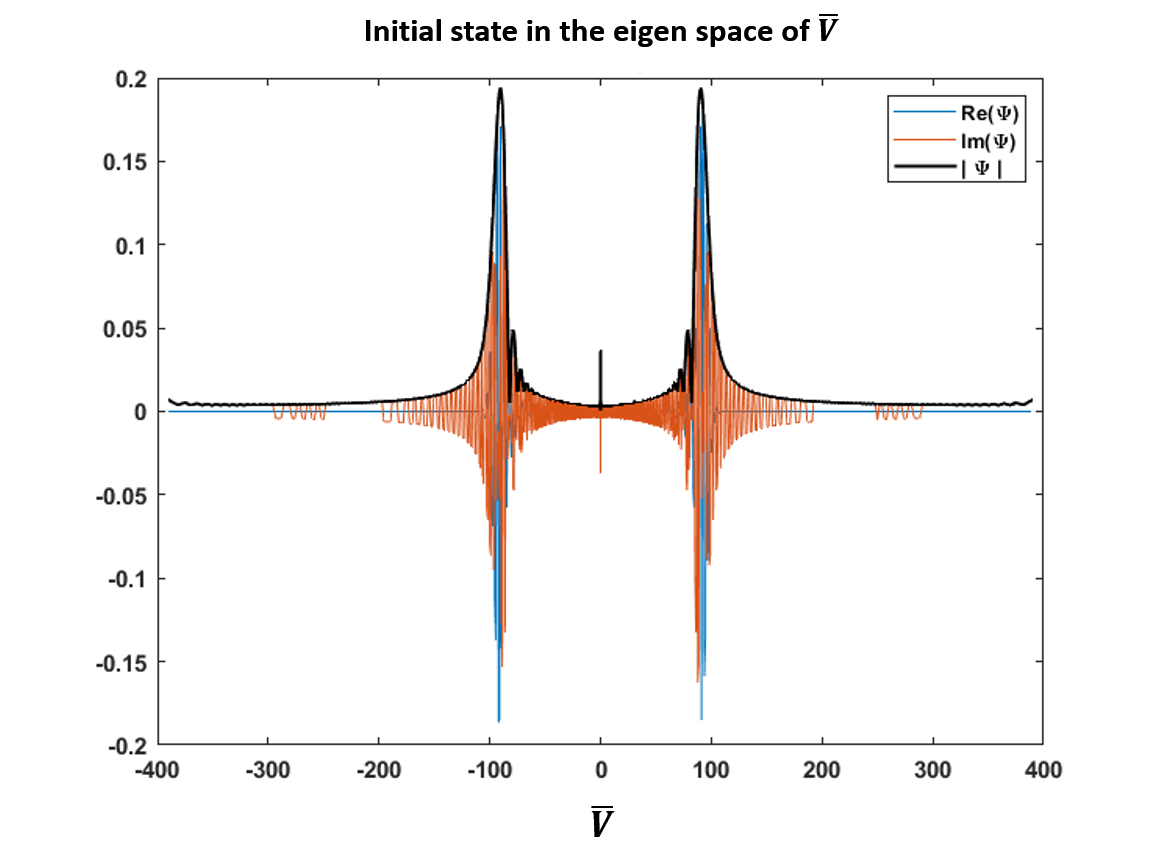}}
\ineps{\includegraphics[width=1\linewidth]{initVbar.eps}}
\inpng{\includegraphics[width=1\linewidth]{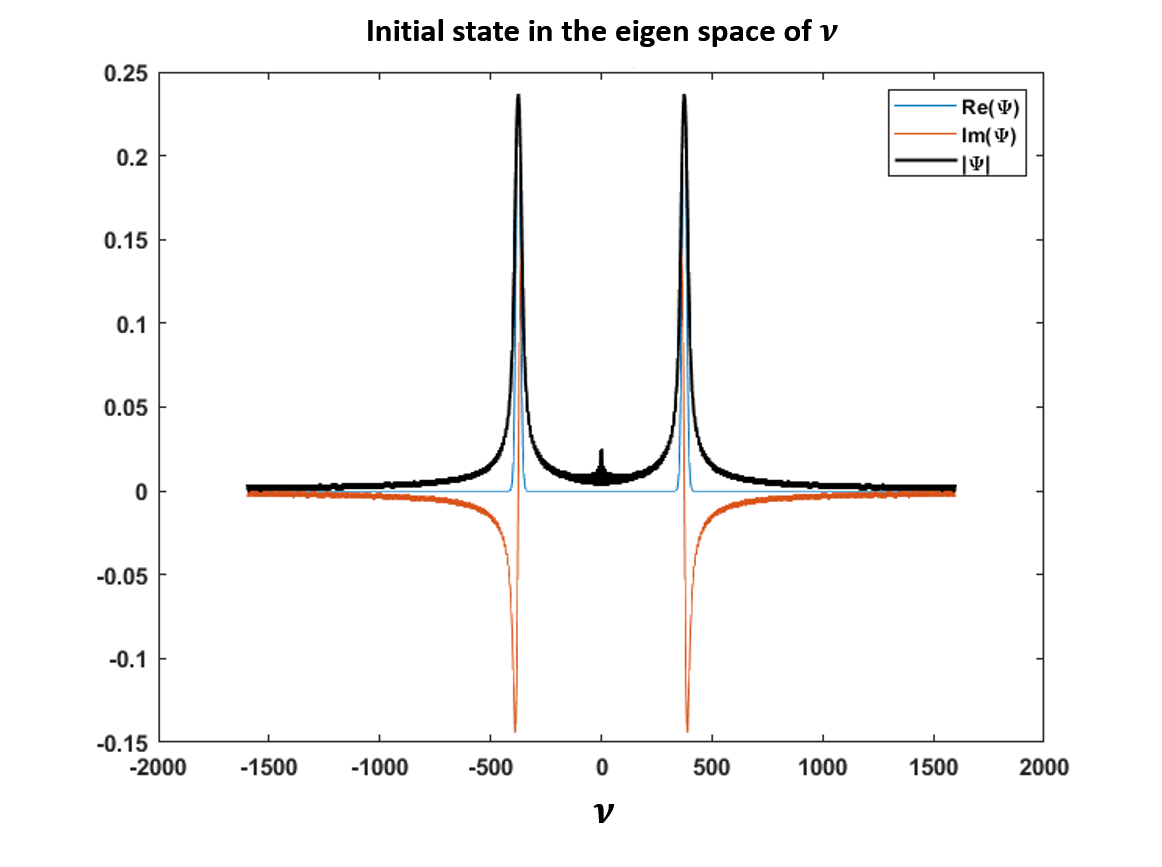}}
\ineps{\includegraphics[width=1\linewidth]{initnu.eps}}
\vspace{-10pt}
\caption{The initial state at $\tau=-7.82$ used in Fig. \ref{fig:Vol} in the Einstein-frame volume $\bar{V}$ basis (upper) and in the $\nu$ basis (lower). This wave packet is chosen to be Gaussian-like localized distribution of corresponding volume states for ease of comparing quantum and classical evolutions.}
\label{fig:Init}
\end{figure}

\begin{figure}
\inpng{\includegraphics[width=0.83\linewidth]{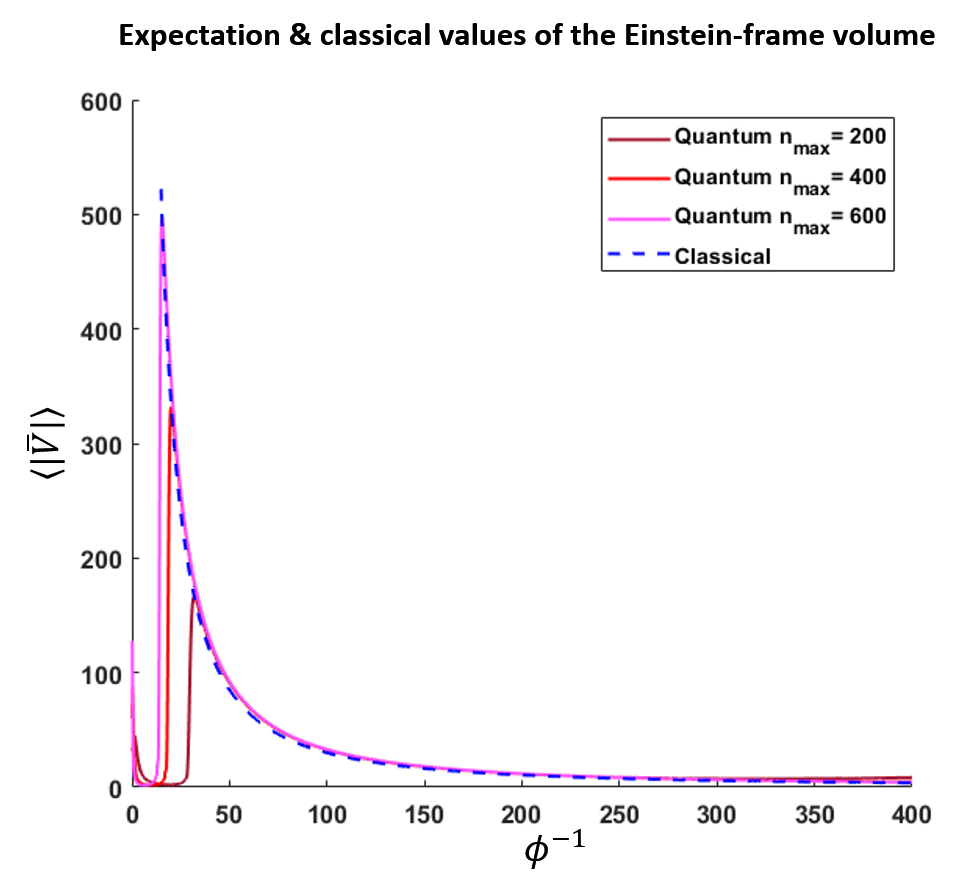}}
\ineps{\includegraphics[width=0.83\linewidth]{AbsVolEin.eps}}
\vspace{-10pt}
\caption{Effects of the grid size on numerical simulations:
The quantum expectation values for the Einstein-frame volume is evaluated using $m=0$ and truncation sizes $n_{\max} =$ 200, 400, and 600, otherwise similar settings  as in Figs. \ref{fig:Vol} and \ref{fig:Init}, including The inverse scalar $\phi^{-1}=e^{\tau/2}$ as the evolution parameter. It can be seen that these expectation values agree with the classical solution (blue dash) except some regions of small $\phi^{-1}$ close to the big bang due to the finite grid size used in the numerical simulations. However, this region narrows with improved grid size using a larger $n_{\max}$.}
\label{fig:AbsVolEin}
\end{figure}

The weakly vanishing of Eq. \eqref{hhhc} gives rise to the Klein-Gordon-like equation
$\bar{\HH}\ket{\Psi}=0$
with
${\Pi} \to i\partial_\tau$
and
${L} \to -i\partial_\theta$.
It admits time-dependent solutions in the separable form
\begin{eqnarray}
\ket{\Psi(\theta,\tau)}
&=&
\sum_{k,m}
\psi_{k m}\,
e^{-i\, \omega(k, m) \tau}
\ket{k,m}
\end{eqnarray}
where
$
\ket{k,m}
=
\frac{1}{\sqrt{2\pi}}\,
e^{i\, 2 m \pi \theta}
\ket{k}
$
in terms of $\ket{k}$ as the orthonormal eigen states of ${\bar{K}}$ with eigen values $k$ which can be derived in a standard manner from the matrix elements of ${\bar{K}}$ using Eq. \eqref{KVHact1},
and
\begin{eqnarray}
\omega^2
&=&
\frac{6\xi-1}{6\xi}\, k^2 + (6\xi-1) \pi^2 m^2
\label{disp}
\end{eqnarray}
for $m=0,\pm1,\pm2, \ldots$


The positive frequency solutions satisfying the Schr\"odinger equation \eqref{scheq2}, i.e.
\begin{eqnarray}
i\partial_\tau
\ket{\Psi}
&=&
H_\text{eff}\,\ket{\Psi}
\label{Sch}
\end{eqnarray}
with the effective Hamiltonian
\begin{eqnarray}
H_\text{eff}
&=&
\sqrt{
\frac{6\xi-1}{6\xi}\,{\bar{K}}^2
+
\frac{6\xi-1}{4}\,{L}^2
}
\label{sqrtham}
\end{eqnarray}
has an inner product
\begin{eqnarray}
\braket{\Psi_1}{\Psi_2}
&=&
\sum_{k_1,k_2,m}
\int_0^{2\pi}\frac{\dd\theta}{2\pi}\,
\psi_{1\,k_1 m}^*\psi_{2\,k_2 m} \braket{k_1}{k_2}
\nppp
&=&
\sum_{k,m}
\psi_{1\,k m}^*\psi_{2\,k m}.
\label{ipd}
\end{eqnarray}
It follows that the basis states $\ket{k,m}$ are orthonormal, satisfying
$\braket{k_1,m_1}{k_2,m_2}=\delta_{k_1 k_2}\delta_{m_1 m_2}$. Furthermore,
the Hamiltonian \eqref{sqrtham} is indeed self-adjoint with respect to the inner product \eqref{ipd} since $H_\text{eff}$ is diagonal in the $\ket{k,m}$ basis with real eigen values $\sqrt{(6\xi-1)(k^2/\,6\xi + \pi^2 m^2)}$.

In the continuous $k$ limit, dispersion relation \eqref{disp} yields a group velocity $v = \dd \bar{U}/\dd\tau$, where $\bar{U} = \ln \bar{V}^{-2/3}$ is canonically conjugate to $\bar{K}$. For example, with $m=0$, we have
\begin{eqnarray*}
v
&=&
\frac{\dd\omega}{\dd k}
=
\pm
\sqrt{
\frac{6\xi-1}{6\xi}
}
\end{eqnarray*}
agreeing with the classical solution
$\bar{U} = \pm\sqrt{
\frac{6\xi-1}{6\xi}
}\,\tau$ up to a translation in $\tau$.

Under a rescaling of the Immirzi parameter
\begin{eqnarray}
\gamma
&\to&
e^\alpha \gamma
\label{gamashft}
\end{eqnarray}
%
we have the induced changes
$b\to e^{-\alpha/2} b$
and
$\nu \to e^{\alpha/2} \nu$
and hence
\begin{eqnarray}
\bar{B}
&\to&
e^{-3\alpha/2} \bar{B}
,\quad
\bar{V}
\to
e^{3\alpha/2} \bar{V} .
\label{2BBVV}
\end{eqnarray}
Therefore, to accompany Eq. \eqref{gamashft} if we also perform a rescaling of the scalar $\phi \to e^{\alpha/2} \phi$, or equivalently a translation of time
\begin{eqnarray}
\tau
\to
\tau - \alpha
\label{phitaushft}
\end{eqnarray}
then the Jordan-frame volume stays invariant, i.e.
\begin{eqnarray}
V
&\to&
V
\end{eqnarray}
under Eqs. \eqref{gamashft} and \eqref{phitaushft} simultaneously. Consequently, the cosmological evolution is free from the Immirzi ambiguity. The Schr\"odinger equation \eqref{Sch} underlines its unitary quantum evolution without singularities at all time values in $\tau$. The big bang singularity emerges only from compactifying the infinite half axis in the negative $\tau$ direction through Eq. \eqref{etau}, where the universe appears to begin with a zero Jordan-frame volume, as illustrated in Figs. \ref{fig:Vol}--\ref{fig:AbsVolEin}.

\section{Conclusion}

Numerous recent developments in physics and cosmology have pointed to significant importance of scale invariance. Supported by this cumulation of evidence, we have presented a novel approach to quantum gravity providing a joint resolution to the vital problem of time, ambiguity of quantized geometry, and existence of a well-behaved big bang, through the principle of scale gauge symmetry. Using background-independent loop quantization techniques, we have constructed a new formulation of time discretely distributed in space as the direct carrier of quantized energy at the most elementary level. The precise determination of this energy will in general involve detailed gravitational interactions with matter encoded in the scale-invariant Hamiltonian constraint, which offers a new ground for unifying quantum and gravitational theories.


\begin{acknowledgments}
The authors are grateful to Y. Adamu and D. \hbox{Rodrigues} for discussions and to
the Carnegie Trust and Cruickshank Trust in Scotland for financial support.
\end{acknowledgments}


\begin{thebibliography}{99}


\bibitem{Blas2011}
D. Blas, M. Shaposhnikov, and D. Zenh\"ausern,
Scale-invariant alternatives to general relativity,
Phys. Rev. D {\bf 84}, 044001 (2011).

\bibitem{Bars2014}
I. Bars, P. Steinhardt, and N. Turok,
Local conformal symmetry in physics and cosmology,
Phys. Rev. D {\bf 89}, 043515 (2014).

\bibitem{Salvio2015}
K. Kannike, G. H\"utsi, L. Pizza, A. Racioppi, M. Raidal, A. Salvio, and A. Strumia,
Dynamically induced Planck scale and inflation,
J. High Energ. Phys. {\bf 05}, 065 (2015).

\bibitem{Ferreira2016}
P. G. Ferreira, C. T. Hill, and G. G. Ross,
Scale-independent inflation and hierarchy generation,
Phys. Lett. B {\bf 763}, 174 (2016).


\bibitem{Ferreira2017}
P. G. Ferreira, C. T. Hill, and G. G. Ross,
Weyl current, scale-invariant inflation, and Planck scale generation,
Phys. Rev. D {\bf 95}, 043507 (2017).

\bibitem{Ferreira2018}
P. G. Ferreira, C. T. Hill, J. Noller, and G. G. Ross,
Inflation in a scale invariant universe,
Phys. Rev. D {\bf 97}, 123516 (2018).


\bibitem{Veraguth2017}
O. J. Veraguth and C. H.-T. Wang,
Immirzi parameter without Immirzi ambiguity: Conformal loop quantization of scalar-tensor gravity,
Phys. Rev. D {\bf 96}, 084011 (2017).


\bibitem{Wang2018}
C. H.-T. Wang and D. P. F. Rodrigues,
Closing the gaps in quantum space and time: Conformally augmented gauge
structure of gravitation,
Phys. Rev. D {\bf 98}, 124041 (2018).

\bibitem{Shaposhnikov2018}
M. Shaposhnikov and A. Shkerin,
Gravity, scale invariance and the hierarchy problem,
JHEP {\bf 10}, 024 (2018).

\bibitem{Wetterich2019}
C. Wetterich,
Quantum scale symmetry,
arXiv:1901.04741,
and references therein.


\bibitem{Isham1992}
C. J. Isham,
Canonical quantum gravity and the problem of time,
NATO Sci. Ser. {\bf 409}, 157 (1993).

\bibitem{Rovelli1994}
C. Rovelli and L. Smolin,
The Physical Hamiltonian in Nonperturbative Quantum Gravity,
Phys. Rev. Lett. {\bf 72}, 446 (1994).

\bibitem{Husain2012}
V. Husain and T. Pawlowski,
Time and a Physical Hamiltonian for Quantum Gravity,
Phys. Rev. Lett. {\bf 108}, 141301 (2012).

\bibitem{Thiemann2008}
T. Thiemann,
{\it Modern Canonical Quantum General Relativity}
(Cambridge University Press, Cambridge, England, 2008).

\bibitem{Rovelli2011}
C. Rovelli,
Loop quantum gravity: The first 25 years,
Classical Quantum Gravity {\bf 28}, 153002 (2011).

\bibitem{Ashtekar2017}
{\it Loop Quantum Gravity: The First 30 Years},
edited by A. Ashtekar and J. Pullin
(World Scientific, Singapore, 2017).


\bibitem{Immirzi1997a}
G. Immirzi,
Real and complex connections for canonical gravity,
Classical Quantum Gravity {\bf 14}, L177 (1997).

\bibitem{Perez2017}
A. Perez,
Black holes in loop quantum gravity,
Rep. Prog. Phys. {\bf 80}, 126901 (2017).

\bibitem{Ashtekar2006}
A. Ashtekar, T. Pawlowski, and P. Singh,
Quantum nature of the big bang: Improved dynamics,
Phys. Rev. D {\bf 74}, 084003 (2006).


\bibitem{Ashtekar2008}
A. Ashtekar, A. Corichi, and P. Singh,
Robustness of key features of loop quantum cosmology,
Phys. Rev. D {\bf 77}, 024046 (2008).


\bibitem{Agullo2016}
I. Agullo and P. Singh,
Loop Quantum Cosmology: A brief review,
Chapter in Ref.~\cite{Ashtekar2017}, and references therein.


\bibitem{Bekenstein1980}
J. D. Bekenstein and A. Meisels,
Conformal invariance, microscopic physics, and the nature of gravitation,
Phys. Rev. D {\bf 22}, 1313 (1980).


\bibitem{Wang2005a}
C. H.-T. Wang,
Conformal geometrodynamics: True degrees of freedom in a truly canonical structure,
Phys. Rev. D {\bf 71}, 124026 (2005).

\bibitem{Wang2005b}
C. H.-T. Wang,
Unambiguous spin-gauge formulation of canonical general relativity with conformorphism invariance,
Phys. Rev. D {\bf 72}, 087501 (2005).


\bibitem{Ashtekar2003}
A. Ashtekar, J. Lewandowski and H. Sahlmann,
Polymer and Fock representations for a scalar field,
Classical Quantum Gravity {\bf 20}, L11 (2003).

\bibitem{Esposito2001}
G. Esposito-Far\`ese and D. Polarski,
Scalar-tensor gravity in an accelerating universe,
Phys. Rev. D {\bf 63}, 063504 (2001).


\bibitem{Achour2017}
J. B. Achour and E. R. Livine,
Thiemann complexifier in classical and quantum FLRW cosmology,
Phys. Rev. D {\bf 96}, 066025 (2017).

\bibitem{Achour2019}
J. B. Achour and E. R. Livine,
Polymer quantum cosmology: Lifting quantization ambiguities using a SL(2,R) conformal symmetry,
Phys. Rev. D {\bf 99}, 126013 (2019).

\bibitem{Wagoner1970}
R. V. Wagoner,
Scalar-tensor theory and gravitational waves,
Phys. Rev. D {\bf 1}, 3209 (1970).






\end{thebibliography}
\end{document}